\documentclass[useAMS,usenatbib,usegraphicx,referee]{mn2e}

\title[HD169142]{The circumstellar disc around the Herbig
AeBe star HD169142}
\author[ ]{W. R. F. Dent$^{1}$, J. M. Torrelles$^{2,3}$, 
M. Osorio$^{4}$, N. Calvet$^{5}$, G. 
Anglada$^{4}$\\
$^{1}$UK Astronomy Technology Centre, Royal Observatory Edinburgh,
Blackford Hill,
Edinburgh EH9 3HJ, UK\\
$^{2}$ Instituto de Ciencias del Espacio (CSIC)-IEEC, C/ Gran Capit\`a, 2-4,
08034 Barcelona, Spain\\
$^{3}$ On sabbatical leave at the UK Astronomy Technology 
Centre, Royal Observatory Edinburgh, UK\\
$^{4}$Instituto de Astrof\'{\i}sica de Andaluc\'{\i}a (CSIC), Apdo. Postal 3004,
18080 Granada, Spain\\
$^{5}$Dept. of Astronomy, University of Michigan, 825 Dennison
Building,
500 Church St., Ann Arbor, MI 48109, USA}
\begin{document}

\date{Received ...; Accepted ...}

\pagerange{\pageref{firstpage}--\pageref{lastpage}} \pubyear{2005}

\maketitle

\label{firstpage}

\begin{abstract}

We present 7~mm and 3.5~cm wavelength continuum observations toward the
Herbig AeBe star HD169142 performed with the Very Large Array (VLA) with
an angular resolution of $\simeq$ 1$''$. We find that this object exhibits
strong ($\simeq$ 4.4 mJy), unresolved ($\la$ 1$''$) 7~mm continuum
emission, being one of the brightest isolated 
Herbig AeBe stars ever detected with
the VLA at this wavelength. No emission is detected at 3.5 cm continuum,
with a 3$\sigma$ upper limit of $\simeq$ 0.08~mJy. From these values, we
obtain a spectral index $\alpha$ $\ga$ 2.5 in the 3.5~cm to 7~mm
wavelength range, indicating that the observed flux density at 7~mm is
most likely dominated by thermal dust emission coming from a circumstellar
disc. We use available photometric data from the literature to model the
spectral energy distribution (SED) of this object from radio to
near-ultraviolet frequencies.  The observed SED can be understood in terms
of an irradiated accretion disc with low mass accretion rate, $\dot M_{\rm
acc}\simeq10^{-8}~M_\odot$~yr$^{-1}$, surrounding a star with an age of
$\simeq$ 10 Myr. We infer that the mass of the disc is 
$\simeq$ 0.04~M$_{\odot}$, and is populated by dust grains that have
grown to a maximum size of 1 mm everywhere, consistent with the lack of
silicate 10 $\mu$m emission. These features, as well as indications of
settling in the wall at the dust destruction radius, led us to speculate
the disc of HD169142 is in an advanced stage of dust evolution, particularly
in its inner regions. 

\end{abstract}


\section{Introduction}

Herbig Ae/Be (HAeBe) stars are young stellar objects (YSOs) of
intermediate mass, characterized by large infrared excesses due to the
presence of circumstellar discs, and are believed to be the more massive
analogues of T Tauri stars (Herbig 1960, Strom et al. 1972). The presence
of discs around isolated HAeBe stars has been mainly inferred by modeling their
spectral energy distribution (SED) at millimetre, submillimetre, infrared,
and optical wavelengths (e.g., Dullemond, Dominik \& Natta 2001; Natta et
al. 2001, 2004; Meeus et al. 2001;  Dominik et al. 2003; Dullemond \&
Dominik 2004; Acke, van den Ancker \& Dullemond 2005;  Hern\'andez et al.
2005). However only a few of these dust/gas structures have been
(marginally) spatially resolved (sizes $\leq$ 2$\times$beam; eg., Mannings
\& Sargent 1997). 

As a continuation of a survey program to study the circumstellar dust/gas
around isolated HAeBe stars (including submillimetre wavelength CO
observations) performed by Dent, Greaves \& Coulson (2005), we have
searched at the VLA archive for centimetre/millimetre wavelength continuum
observations toward those stars where submillimetre CO emission was
detected. The main goal of that search was to identify dust disc
candidates which may be resolvable through future high angular continuum
observations. By reducing and analyzing the available VLA archive data on
these sources, we have found in particular that the HAeBe star
HD169142 is associated with strong 7~mm continuum emission (data presented
in this paper). 

The A5Ve star HD169142 is a relatively
nearby ($d$ = 145 pc) example of an isolated HAeBe star (Dunkin,
Barlow \& Ryan 1997, van den Ancker 1999, Meeus et al.
2001). It has significant far-infrared emission as well as a mid-infrared
excess (Malfait, Bogaert \& Waelkens 1998), and is one of the brightest
such objects at submillimetre wavelengths (Sylvester et al. 1996).
The absence of large-scale nebulosity or extended
molecular gas in the region suggests no nearby on-going star formation and that
HD169142 is a relatively evolved HAeBe star.  HD169142 also shows a bright and
narrow submillimetre emission CO line (Greaves, Mannings \& 
Holland 2000; Dent, Greaves
\& Coulson 2005), as well as a narrow optical OI line profile (Acke, van
den Ancker \& Dullemond 2005). These results indicate that any
circumstellar disc around the star must be close to face-on, with an
inclination angle of $\simeq$ 10$^{\circ}$ (Dent, Greaves \& Coulson
2005). Attempts to resolve the continuum emission have
resulted in an upper limit to the mid-infrared disc radius of 150 Astronomical
Units (AU) (Jayawardhana et al. 2001), although there is evidence of
polarized near-infrared emission extending to $\simeq$ 200 AU (Kuhn,
Potter \& Parise 2001). An unpublished Submillimetre Common-User
Bolometer Array (SCUBA) map from the James Clerk Maxwell Telescope (JCMT)
archive shows unresolved emission at 850 and 450 $\mu$m, with a
deconvolved upper limit to the size of $\simeq$ 4$''$ (FWHM) (or a radius of
$\simeq$ 300 AU) in the dust continuum at 450 $\mu$m. Finally, observations
of the 3.3 $\mu$m feature of Polycyclic Aromatic Hydrocarbons (PAHs) show
that this emission arises from an extended region, with a size of
$\simeq$ $0\farcs3$, or 43~AU (Habart, Natta \& Krugel 2004). 

In this paper we present 3.5 cm and 7~mm continuum observations carried
out with the VLA toward the source HD169142. We discuss in \S~2.1 the
nature of its continuum emission at cm/mm wavelengths. In addition to the
VLA data, in \S~2.2 we compile data from the literature to construct the
SED of HD169142 over a wide range of wavelengths, from radio to
near-ultraviolet.  In \S~3 we compare the SED with a
grid of self-consistent irradiated accretion disc models from D'Alessio et al.
(2005).  Our studies imply that a disc with mass $\simeq$
0.04~M$_{\odot}$
and a population of grains grown to a maximum
radius of $\simeq$ 1~mm surrounding a star with an age of 10 Myr can
reasonably explain the main characteristics of the observed SED. 

\section[]{Observations}

\subsection[]{VLA observations and results}

The 3.5 cm and 7 mm wavelength continuum observations were carried out
with the VLA of the National Radio Astronomy Observatory
(NRAO)\footnote[6]{The NRAO is a facility of the National Science
Foundation operated under cooperative agreement by Associated
Universities, Inc.} in the B (2005 May 15) and CnB (2002 September 22-23)
configurations, respectively. For both wavelengths, a bandwidth of 100 MHz
and two circular polarizations were used. All data were reduced with the
Astronomical Image Processing System (AIPS) of NRAO using standard VLA
procedures.  The absolute amplitude calibrator was 3C286, with assumed
flux densities of 5.2 Jy ($\lambda$=3.5~cm) and 1.5 Jy ($\lambda$=7~mm),
while the phase calibrators were 1911$-$201 (for the $\lambda$=3.5 cm
observations, with a bootstrapped flux density of 2.4~Jy) and 18210$-$25282
(for the $\lambda$=7~mm observations, with a bootstrapped flux density of
0.6~Jy). Cleaned maps were made by Fourier transforming the ($u$,$v$) data
with natural weighting using the IMAGR task of AIPS. The resulting
synthesized beam sizes and rms sensitivities of the maps were $\simeq$
$1\farcs7\times 0\farcs7$ (PA = 8$^{\circ}$) and $\simeq$
25~$\mu$Jy~beam$^{-1}$ at 3.5~cm wavelength, and $\simeq$ $0\farcs9\times
0\farcs 4$ (PA = $-35^{\circ}$) and $\simeq$ 300~$\mu$Jy~beam$^{-1}$ at
7~mm wavelength, respectively (Table 1). 

\begin{table*}
\begin{minipage}{\textwidth}
\caption{\sf Summary of the VLA observations of HD169142}
\begin{tabular}[t]{ccccc@{\extracolsep{0.5em}}cccc} \hline\
$\lambda$  & Date & Array & \multicolumn{2}{c}{Phase Calibrator} & \multicolumn{2}{c}{Beam$^b$} 
& rms$^b$  & HD169142 \\
  \cline{4-5} \cline{6-7}
  (cm)  &           &  Configuration & Name & $F_\nu$$^a$ (Jy) & Size & PA & ($\mu$Jy beam$^{-1}$) & $F_\nu$$^c$ (mJy) \\
\hline
0.7 & 2002 Sep 22-23 & CnB & 18210$-$25282 & 0.6 & 
$0\farcs9 \times 0\farcs4$ & $-$35$^{\circ}$  & 300 & 4.4\\
3.5  & 2005 May 15 & B & 1911$-$201 & 2.4 &
$1\farcs7\times 0\farcs7$ & $+$8$^{\circ}$ & 25 & $\la$ 0.08$^d$ \\
\hline
\end{tabular}

$^a$Bootstrapped flux density of the phase calibrator.\\
$^b$For natural weight maps.\\
$^c$Flux density of the source.\\
$^d$3$\sigma$ upper limit.
\end{minipage}
\end{table*}

A bright ($\simeq$ 4.4~mJy~beam$^{-1}$), unresolved ($\la$ 1$''$) 7~mm
continuum source is detected toward HD169142 (see Figure 1), with its peak
position at $\alpha({\rm J2000})$ = $18^h$ $24^m$ $29\fs 78$, $\delta({\rm
J2000})$ = $-29^\circ$ $46'$ $49\farcs 4$ ($\pm0\farcs1$), coinciding with
the optical position of the star (Hog et al. 1998). To our knowledge, this
is one of the brightest 7 mm continuum sources ever detected with the VLA
toward a HAeBe star (from the literature we see that only HD163296, with a
flux density of 6 mJy at 7 mm, is brighter; Natta et al. 2004). 

\begin{figure}
\hspace{-30pt}
\includegraphics{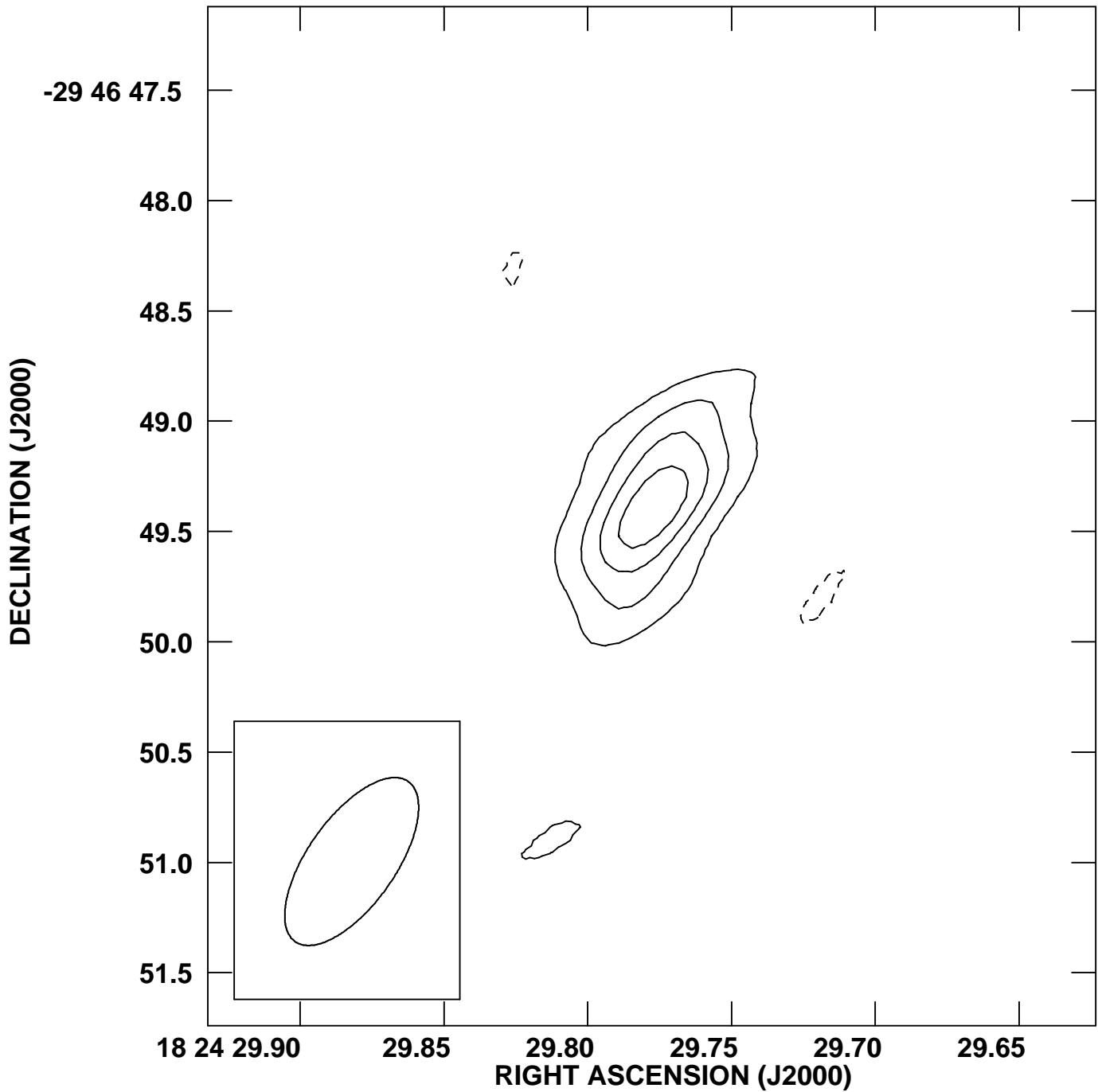} 
 \caption{Contour map of the $\lambda=$7~mm continuum emission of HD169142
observed with the VLA. Contour levels are $-$3, 3, 6, 9, and
12$\times$0.3~mJy~beam$^{-1}$, the rms of the map. The beam
(size=$0\farcs9\times 0\farcs 4$, PA = $-35^{\circ}$) is shown in the
lower left corner of the map. This bright ($\simeq$ 4.4~mJy~beam$^{-1}$)
unresolved ($\la$ 1$''$) 7~mm continuum source is most likely originated
from dust emission in a circumstellar disc around HD169142 (see \S 3).}
 \end{figure}

No 3.5 cm continuum emission is detected toward HD163296, with an upper
limit of $\simeq$ 0.08~mJy~beam$^{-1}$ (3$\sigma$ level). From these
values, we obtain a spectral index $\alpha$ $\ga$ 2.5 ($F_{\nu} \propto
\nu^{\alpha}$) in the 3.5~cm to 7 mm wavelength range.  This high value of
the spectral index indicates that free-free emission cannot account for
the observed flux density at 7~mm, and that thermal dust emission from a
circumstellar disc is most likely the main contribution to the observed
flux density at this wavelength (see also \S~3). 

\subsection[]{Spectral Energy Distribution}

In Table 2, we list the flux density values of HD169142 from
near-ultraviolet to radio wavelengths used for modeling the SED (\S~3). 
The magnitudes UBV and JHK measured by Malfait, Bogaert \& Waelkens (1998)
and Sylvester et al. (1996) have been dereddened using a value of the
extinction to the star of $A_{\rm V} \simeq $ 0.5, which is obtained from
the standard extinction law with $R_{\rm V} \simeq$ 3.1 and a spectral
type A5.  Magnitudes have been converted into flux densities using zero
points from Johnson (1966) and Bessell \& Brett (1988). We also use
Infrared Space Observatory (ISO) Short Wavelength Spectrometer (SWS) data
obtained toward HD169142 (Meeus et al. 2001).

\begin{table*}
\begin{minipage}{80mm}
\caption{\sf Flux density values of HD169142$^a$}
\begin{tabular}[t]{llc} \hline\
$\lambda$   & $F_\nu$$^b$ & Reference \\
 ($\mu$m)  &      (Jy) & \\
\hline
0.36(U)      &      2.0                  & 1 \\
0.45(B)      &      5.38                 & 2 \\
0.55(V)      &      4.56                 & 2 \\
0.90(I)      &      1.79                 & 2 \\
1.22(J)      &      2.0                  & 2 \\
1.65(H)      &      1.78                 & 2 \\
2.20(K)      &      1.58                 & 2 \\
3.45(L)      &      1.61                 & 1 \\
3.80(L')     &      1.29                 & 1 \\
4.8(M)       &      0.96                 & 1 \\
10.8(N)      &      2.37$\pm$0.2         & 3 \\
12           &      2.95$\pm$0.3         & 4 \\
18.2(IHW)    &      7.86$\pm$0.8         & 3 \\
25           &      18.4$\pm$1.8         & 4 \\
60           &      29.6$\pm$3.0         & 4 \\
100          &      23.4$\pm$2.3         & 4 \\
450          &      3.34$\pm$0.115       & 5 \\
800          &      0.554$\pm$0.034      & 2 \\
850          &      0.565$\pm$0.1        & 5 \\
1100         &      0.287$\pm$0.013      & 2 \\
1300         &      0.197$\pm$0.015      & 2 \\
2000         &      0.070$\pm$0.019      & 2 \\
7000         &      0.0044$\pm$0.0009    & 6 \\
35000        &      $\leq$ 0.00008       & 6 \\
\hline
\end{tabular}

References: (1) Malfait, Bogaert \& Waelkens (1998); (2) 
Sylvester et al. (1996); (3) Jayawardhana et
al. (2001); (4) IRAS PSC; (5) Sandell \& Weintraub (2005); (6) This
work.\\
$^a$In addition to the values listed in this table we also use 
ISO short-wavelength spectrometer data obtained toward HD169142 by 
Meeus et al. (2001).\\
$^b$ Uncertainties in the near-ultraviolet through near-infrared flux densities
are less than 10$\%$.
\end{minipage}
\end{table*}

\section{Discussion and conclusions}

HD169142 is considered a relatively evolved Herbig Ae/Be star with an A5Ve
spectral type (Dunkin, Barlow \& Ryan 1997). Therefore, its circumstellar
envelope is likely gone and only the disc is responsible for the dust
emission at all wavelengths. At present, HD169142 is one of the few
intermediate-mass stars where emission at wavelengths as long as 7 mm has
been detected.  Consequently, the good wavelength coverage of the SED of
HD169142, with observational data ranging from near-ultraviolet to radio
frequencies (Table 2 and Figure 2), makes this object a good candidate to
find a self-consistent irradiated accretion disc model, with parameters
reasonably well constrained.

\begin{figure}
\hspace{-50pt}
\includegraphics{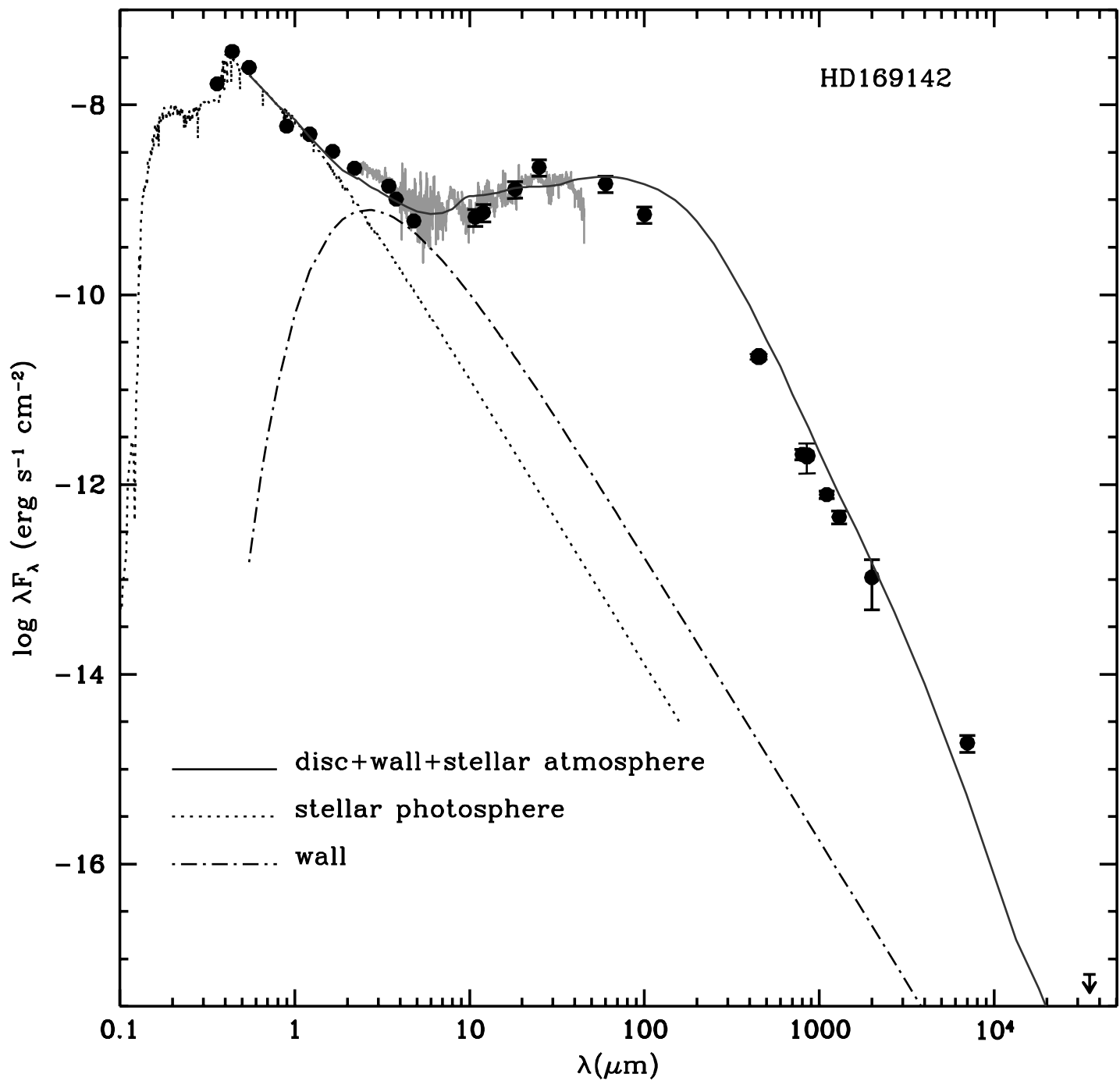} 
 \caption{Observed and model SEDs for the source HD169142.  The dotted
line corresponds to the emission from the stellar photosphere of an A2
star with an age of 10 Myr.  The dotted-dashed line is the emission
arising from the inner cylindrical surface where the disc is truncated by
dust sublimation (the ``wall'' with a height of 0.018 AU).  The solid-line
represents the total emission that includes the disc (whose physical
parameters are $i=30^{\circ}$, ${\dot M}_{\rm acc}=10^{-8}$ $M_{\odot}$
yr$^{-1}$, $R_{\rm disc} \simeq$ 300 AU and $a_{\rm max}=1$ mm), the
stellar photosphere and the ``wall'' contributions (see \S~3 and Table 3).
Thin gray lines indicate ISO short-wavelength spectrometer values obtained
by Meeus et al. (2001).}
 \end{figure}

The observed SED (Figure 2) exhibits three remarkable characteristics:
strong millimetre emission, a shallow dependence of the long wavelength
dust opacity on wavelength ($\kappa \propto \lambda^{-0.5}$, as indicated
by the spectral index in the submillimetre and millimetre ranges, assuming
the total flux density 
is dominated by optically thin emission), and absence of a
detectable silicate feature at 10 $\mu$m (Meeus et al. 2001). To account
for the observed strong millimetre emission, either a massive disc or a
dust mixture with large grains is required. If the disc is assumed to have
interstellar-like grains, characterized by a small emissivity at
millimetre wavelengths, then it is required a disc more massive than if
the grains have a maximum size around 1 mm (D'Alessio, Calvet \& Hartmann
2001); in this last case, the grains have the largest possible emissivity
at millimetre wavelengths and then the required mass of the disc is the
minimum possible. In an accretion $\alpha$-disc irradiated by the central
star, the surface density $\Sigma$ scales as $\dot M_{\rm acc}/\alpha$
(D'Alessio et al. 1999), where $\dot M_{\rm acc}$ is the disc mass
accretion rate and $\alpha$ is the viscosity parameter (Shakura \& Sunyaev
1973). Therefore, for a given disc radius, the increase in the disc mass
resulting from increasing $\dot M_{\rm acc}$ is equivalent to that
produced by decreasing $\alpha$ by the same amount. If the value of the
disc radius is constrained from observations, a massive disc can be
obtained either with high values of $\dot M_{\rm acc}$
($>10^{-6}~M_\odot$~yr$^{-1}$), or by assuming small values of $\alpha$.
However a massive disc would probably imply a mid-infrared emission higher
than observed. A mixture including large grains is also favoured instead
of a massive disc, because of the shallow dependence of the dust opacity
with wavelength in the submillimetre and millimetre ranges (implying
$\beta=0.5$, where $\kappa \propto \lambda^{-\beta}$, while smaller grains
have typically $\beta$ = 1.5-2; see for example the dependence of the
millimetre flux densities and slope on the grain size in Figure 8 of
Osorio et al. 2003), the lack of silicate emission (see Fig. 1 in
D'Alessio, Calvet \& Hartmann 2001), and the fact that massive discs may
become gravitationally unstable (see below).

In order to derive the physical parameters of HD169142, we compare the
observed SED with the database of structural parameters and synthetic SEDs
of irradiated disc models published by D'Alessio et al.
(2005)\footnote[7]{http://www.astrosmo.unam.mx/$\sim$dalessio/}. The SEDs
in this catalog include the contribution of an irradiated accretion disc,
the contribution of the stellar photosphere (relevant in the optical
regime) calculated using Kurucz (1993) models, and the contribution of the
inner cylindrical surface that faces the star, where the dust disc is
truncated by dust sublimation at $T \simeq 1400$ K (i.e., the ``wall" at
the dust destruction radius, $R_{\rm wall}$, whose emission is important
in the near-IR range; Natta et al. 2001).

The main characteristics of these disc models are described in D'Alessio
et al. (1998, 1999, and 2001). In summary, these discs are assumed to be
in steady state (with a constant ${\dot M_{acc}}$) and geometrically thin
(i.e., the gas scale height along the disc is smaller than the radial
distance). They are heated by stellar radiation, viscous dissipation, and
ionization by energetic particles. The viscosity parameter $\alpha$ is
calculated following the prescription from Shakura \& Sunyaev (1973),
where the usual value $\alpha=0.01$ is adopted for all the models in the
database, except for the models with 4000 K and 1 Myr where values
$\alpha$ = 0.1, 0.01, and 0.001 were used. The opacity is mainly due to
dust, whose grains are assumed to have a size distribution $n(a) \propto
a^{-3.5}$ between a minimum size $a_{\rm min}$ (fixed at 0.005 $\mu$m) and
a maximum size $a_{\rm max}$ (taken as a free parameter).  In the models
of D'Alessio and collaborators the vertical structure and emission
properties of the disc are calculated self-consistently with the stellar
parameters, instead of using simple power-law descriptions for the dust
temperature and the surface density.

In the case of HD169142, we have a priori constraints for some of the
parameters. For instance, the spectral type of the star is A5Ve (Dunkin,
Barlow \& Ryan 1997), the disc radius should be of the order of 100 AU 
(a value suggested by the mid and near-infrared wavelength
observations,
see \S 1, as well as the 7~mm continuum observations presented in this 
paper), and the inclination angle of the disc
with respect to the line-of-sight is $\simeq 10^{\circ}$ (constrained by
CO observations carried out by Dent, Greaves \& Coulson 2005).  Thus, we
have searched among the models of the D'Alessio et al. (2005) database
those having this set of parameters closer to the actual values of
HD169142. In this way, we have selected an A2 star (whose mass
$M_*=2~M_{\odot}$ corresponds to a stellar luminosity $L_*= 17~L_{\odot}$
and radius $R_*=1.7~R_{\odot}$, according to pre-main sequence tracks from
Siess, Dufour \& Forestini 2000), a disc radius $R_{\rm disc} \simeq$ 300
AU, and an inclination angle $i \simeq 30^{\circ}$.  We have estimated the
values of the remaining parameters by fitting the observed SED. We obtain
a reasonable fit by assuming a central star with an age of 10~Myr,
surrounded by a disc with a mass accretion rate $\dot M_{\rm acc} \simeq
10^{-8} ~M_\odot$~yr$^{-1}$, a mass $M_{\rm disc} \simeq 0.04~M_{\odot}$
(calculated by integrating the surface density in the disc, $\Sigma$, from
the dust destruction radius, $R_{\rm wall}=0.35$ AU, to the outer radius,
$R_{\rm disc}=300$ AU), and where dust grains have grown up to millimetre
sizes ($a_{\rm max}=1$ mm).  We have modified the nominal value of the
height of the ``wall'' at the dust destruction radius used in the
database, in order to obtain a better fit to the observed near-infrared
emission. We obtained the best fit for a height of the dust wall $z_{\rm
wall}=0.018$ AU.  The gas scale height at $R_{\rm wall}$ can be calculated
assuming that there the disc is vertically isothermal with a temperature
of 1400 K. Using equation (13) of Dullemond et al. (2001) we obtain a
value of $\sim$0.01 AU for the gas scale height, implying a ratio of $\sim$2
between the height of the dust wall and the gas scale height. This low
value of the ratio between the height of the dust wall and the gas scale
height (as compared with the value of $\sim$4 usually found in HAeBe
stars; Dullemond et al. 2001) may imply that the dust in the inner disc
has grown and is settling towards the equatorial disc plane, consistent
with the large grains required to explain the 7~mm continuum emission.
Figure 2 shows the total emission of this model (solid-line) that includes
the disc contribution, the stellar photospheric contribution (dotted-line)
and the ``wall'' emission (dotted-dashed line).

We estimate the accretion luminosity as $L_{\rm acc} \simeq G M_* {\dot
M}/R_{*} = 0.37~L_{\odot}$, which is considerably lower than the derived
stellar luminosity ($L_*= 17~L_\odot$); irradiation is therefore the main
heating agent of the disc upper layers.  We note that some regions of the
disc could be unstable to axisymmetric gravitational perturbations if the
Toomre $Q$-parameter ($Q=(c_{\rm s}/\pi \Sigma) (M_*/G R^3)^{0.5}$, where
$c_{\rm s}$ is the sound speed) is less than unity (Toomre 1964). Since
the most unstable regions are expected to occur at large radii, we
evaluated the Toomre $Q$-parameter at the disc radius ($R_{\rm disc}$ =
300 AU, where the surface density is 0.6 g cm$^{-2}$ and the mid-plane
temperature is 20 K, implying $c_{\rm s}$ = 0.3~km~s$^{-1}$) and we
obtained a value of $Q \simeq 13$, implying that the disc is
gravitationally stable. A disc $\sim$10 times more massive will be
$\sim$10 times denser, and would start to become gravitationally unstable.  
A summary of the main parameters of the circumstellar disc around HD169142
is given in Table 3.

\begin{table*}
\begin{minipage}{\textwidth}
\caption{\sf Adopted parameters of the circumstellar disc around HD169142$^a$}
\begin{tabular}[t]{ccccccccc} \hline
Age  & $\dot M_{\rm acc}$ & $L_{\rm acc}$ & $R_{\rm wall}$ & $R_{\rm 
disc}$ & $M_{\rm disc}$ & $i$ & $a_{\rm max}$ & $z_{\rm wall}$ \\
(Myr)  & ($M_\odot$~yr$^{-1}$) & ($L_\odot$) & (AU) & (AU) & ($M_\odot$) 
& (deg) &
(mm) & (AU)\\
\hline
10     & 10$^{-8}$  & 0.37 & 0.35  & 300 & 0.04 & 30 & 1 & 0.018 
\\
\hline
\end{tabular}

$^a$Obtained by modeling the observed SED (see \S~3).
\end{minipage}
\end{table*}

The SED of HD169142 has been modeled by Dominik et al. (2003) using a
passive disc and dust grains with size $\simeq 0.1~\mu$m.  To fit the SED,
Dominik et al. used a power-law radial distribution of surface density,
with a very steep dependence, $R^{-2}$; in contrast, in the model outlined
above the surface density is calculated self-consistently assuming an
$\alpha$-disc model and goes to a $R^{-1}$ dependence at large radii
(D'Alessio et al. 1999). In addition, Dominik et al. had to adopt a
$\lambda^{-1}$ dependence for the opacity in the millimetre range, which
in our case arises naturally from the grain size distribution. Moreover,
their predicted SED had significant silicate emission, while the large
value of the grain maximum size of our adopted dust mixture consistently
predict no emission in this feature. In fact, the large grain sizes, lack
of silicate emission, and the low height of the wall at the dust
destruction radius, all seem to suggest a large degree of grain
coagulation and settling has occured in the inner disc of HD169142. This
is generally consistent with the predictions from models of the
distribution of solids in proto-planetary discs (Weidenschilling 1997;
Dullemond \& Dominik 2004, 2005). Grains are shown to grow and settle very
rapidly, possibly leaving a disc atmosphere dominated by smaller dust. The
small grains can be replenished through mutual collisions. However, in the
case of HD169142, the absence of silicate feature suggests that few grains
smaller than $\sim$~3$\mu$m exist in the disc atmosphere. The maintenance
of relatively large grains in the atmosphere would require an efficient
stirring mechanism; alternatively the dust could be depleted in silicate
for some reason. Why the mm-sized grains have not settled further in such
an old disc is unclear. However, the results of Dullemond \& Dominik
(2004) suggest that it is possible for a thin disc of relatively large
grains to reach a semi-equilibrium state, given a suitably efficient
stirring mechanism. 

Finally, it is important to mention that the current model for HD169142
may not be unique, and that a model with a lower accretion rate (${\dot
M_{\rm acc}}\simeq10^{-9}~M_{\odot}$ yr$^{-1}$) and a smaller value of the viscosity parameter ($\alpha \simeq 0.001$) might be also feasible.  In
this sense, we also note that the slope of the model SED in the far-IR to
submm range is somewhat steeper than observed, suggesting that the
power-law index of the dust grain size distribution is shallower than the
value of $-$3.5 adopted here. Nevertheless, as pointed out by D'Alessio et
al. (2005), to find an unique model, additional observational constraints
such as spatial distribution profiles at various wavelengths are needed.
In fact, the strong 7~mm dust continuum emission reported in this paper
makes this source an excellent candidate to resolve its dust disc contents
both at that particular wavelength with the VLA in its A configuration
(angular resolution $\simeq$ 0\farcs1), as well as with the Expanded
Submillimeter Array (eSMA = JCMT+SMA) at submillimetre wavelengths.

\section*{Acknowledgments}

We deeply thank the referee for her/his very stimulating comments and 
suggestions, that have been most useful to improve the scientific
content of this paper. G.A., M.O., and J.M.T. are supported in part by 
Spanish AYA2005-08523 grant. G.A. and M.O. acknowledge partial support
from Junta de Andaluc\'{\i}a, Spain. M.O. acknowledges support from IAU
Peter Gruber Foundation and MAE-AECI. We thank Gwendolyn Meeus for
providing us the original {\it ISO} data.

\label{lastpage}

\end{document}